\documentclass[12pt,a4paper]{article}
\usepackage{amssymb}
\usepackage{amsfonts}
\usepackage{amsmath}
\usepackage{latexsym}
\usepackage{epsf,epsfig,rotating}
\usepackage{url}
\usepackage{multicol}
\usepackage{cite}

\begin{document}

\title{\begin{flushright}
{\small INR-TH-2016-012}
\end{flushright} {\bf Q-balls in the Wick--Cutkosky model}}

\author{
 E.\,Ya.\;Nugaev$^{a,b}$\thanks{{\bf e-mail}: emin@ms2.inr.ac.ru}
, M.\,N.\;Smolyakov$^{c,a}$\thanks{{\bf e-mail}: smolyakov@theory.sinp.msu.ru}
\\
$^a${\small{\em
Institute for Nuclear Research of the Russian Academy
of Sciences,}}\\
{\small{\em 60th October Anniversary prospect 7a, Moscow 117312,
Russia
}}\\
$^b${\small{\em Moscow Institute of Physics and Technology,
}}\\
{\small{\em Institutskii per. 9, Dolgoprudny, Moscow Region 141700,
Russia
}}\\
$^c${\small{\em Skobeltsyn Institute of Nuclear Physics, Lomonosov Moscow
State University,
}}\\
{\small{\em Moscow 119991, Russia}}}

\date{}
\maketitle

\begin{abstract}
In the present paper Q-ball solutions in the Wick--Cutkosky model are examined in detail. A remarkable feature of the Wick--Cutkosky model is that it admits analytical treatment for the most part of the analysis of Q-balls, which allows one to use this simple model to demonstrate some peculiar properties of Q-balls. In particular, a method for estimating the binding energy of a Q-ball is proposed. This method is tested on the Wick--Cutkosky model taking into account the well-known results obtained for this model earlier.
\end{abstract}

\section{Introduction}
Non-topological solitons in the scalar field theory, which were initially proposed in \cite{Rosen0} and now are known as Q-balls \cite{Coleman:1985ki}, are widely discussed in the literature. However, among the variety of models providing Q-ball-type solutions, only a few of them admit of analytical treatment in four-dimensional space-time, at least for examining their main properties such as the energy--charge dependencies. The existence of analytical solutions simplifies the analysis considerably and allows one to perform a deeper study of the Q-ball properties. For such exceptions, one can recall the model with a very simple polynomial potential proposed in \cite{Anderson:1970et}, where the energy--charge dependence can be obtained analytically, as well as the models of \cite{Rosen1,Arodz:2008jk,Theodorakis:2000bz,Gulamov:2013ema} providing exact analytical solutions for Q-balls (the logarithmic scalar field potential of \cite{Rosen1} makes it possible to examine analytically even the linear perturbations above the Q-ball solution; see \cite{MarcVent}).

All the models mentioned above deal with a single complex scalar field. Meanwhile, there is another class of models, namely, with two different scalar fields (the complex one and the real one), which also admit the existence of Q-ball-like solutions. The best-known example of this type was proposed in \cite{Friedberg:1976me}, in which only an approximate analytical Q-ball solution\footnote{Although solutions in two-field models like the one of \cite{Friedberg:1976me} are not Q-balls in the sense of Coleman's definition of Q-balls \cite{Coleman:1985ki}, they are of the same kind, so from now on we call such soliton solutions ``Q-balls''.} can be obtained (using the trial functions), thus being demanding for numerical calculations. A simplification of this model by neglecting the potential of the real scalar field was performed in \cite{Levin:2010gp}. Q-ball solutions in such a theory have a rather interesting property -- they possess a (non-conserved) ``scalar charge'', which characterizes the long-range attraction between such Q-balls. However, exact analytical formulas for the energy--charge dependence cannot be obtained in this case too.

In the present paper we consider an even more simplified two-field model. Instead of the quartic interaction of the scalar fields in \cite{Levin:2010gp}, we consider the triple Yukawa interaction between the scalar fields, resulting in the well-known Wick--Cutkosky model \cite{Wick,Cutkosky}. Surprisingly, most of the analysis in this model can be performed analytically. In particular, the exact form of the corresponding energy--charge dependence can be obtained analytically (which simplifies the use of the well-know stability criteria), whereas the numerical solution is necessary only for obtaining the value of some universal dimensionless parameter (from this point of view the model is similar to the one-field model of \cite{Anderson:1970et}).

Q-balls are classical objects, but it is clear that they can be considered as bound states of scalar particles of the theory. However, it is not clear how to perform a consistent analysis for the case in which such a bound state consists of a large number of particles, which is exactly the case of a Q-ball. Indeed, even the problem of two bound scalar particles in the Wick--Cutkosky model is not trivial, though it can be solved analytically (see, for example, review \cite{Nakanishi:1988hp}). In the present paper we will propose a simple method that will allow one at least to roughly estimate the binding energy of a Q-ball. This method can be applied to Q-balls in different models, but we will test it on Q-balls of the Wick--Cutkosky model. For such an analysis the Wick--Cutkosky model is unique -- on the one hand, there exist Q-ball solutions in this model; on the other hand, the Bethe--Salpeter equation describing the bound state of two massive scalar particles can be solved analytically in the Wick--Cutkosky model. The latter will allow one to compare the results obtained with the help of the methods of classical and quantum field theories.

\section{Setup and equations of motion}
Let us start with the action of the Wick--Cutkosky model, describing the complex scalar field $\chi$ interacting with the real massless scalar field $\phi$ in the flat four-dimensional space-time with the coordinates $x^{\mu}=\{t,\vec x\}$, $\mu=0,1,2,3$; in the form
\begin{equation}
S=\int\left(\partial_\mu\chi^*\partial^\mu\chi+\frac{1}{2}\partial_\mu\phi\partial^\mu\phi-h\phi\chi^*\chi\right)d^4x,
\label{sys}
\end{equation}
where $h\ne 0$ is the coupling constant of the scalar Yukawa interaction.

Classical vacua of the theory correspond to the stationary points of the potential
\begin{equation}
V(\phi,\chi)=h\phi\chi^*\chi.
\label{potential}
\end{equation}
The corresponding vacuum solutions are just $\chi\equiv 0$ and $\phi=\phi_{0}\equiv\textrm{const}$, i.e., there is a flat direction along the real field. Now let us consider the quadratic part of action (\ref{sys}) for the fluctuations $\chi(t,x)$, $\phi(t,x)=\phi_{0}+\rho(t,x)$ above the vacuum solution. We get
\begin{equation}
S_{(2)}=\int\left(\partial_\mu\chi^*\partial^\mu\chi+\frac{1}{2}\partial_\mu\rho\partial^\mu\rho-h\phi_{0}\chi^*\chi\right)d^4x,
\end{equation}
We see that for $h\phi_{0}<0$ the effective mass $m=\sqrt{h\phi_{0}}$ of the field $\chi$ is imaginary, leading to tachyonic instability. For $h\phi_{0}>0$ the corresponding mass term has the proper sign and one expects that such vacua are stable. As for the case $\phi_{0}=0$, it is possible to show analytically that there are no Q-ball-type solutions for $h\phi_{0}\le 0$ (this topic will be discussed below). For these reasons, below we will consider only the case $h\phi_{0}>0$.

The fact that globally the scalar field potential is not bounded from below is not dangerous -- the vacuum solution is classically stable for $h\phi_{0}>0$.\footnote{Moreover, it will be shown below that the scalar field potential can be modified to become bounded from below while keeping in the theory the most interesting Q-ball solutions.} As for the flat direction, its existence is also not dangerous. Indeed, let us
consider a scalar field (not interacting with gauge fields) with the standard ``Mexican hat''-type potential. There is a class of vacuum solutions, all of them having the same zero energy –-- there exists a flat direction. But the existence of the massless Goldstone bosons does not indicate any instability of such a vacuum, because an infinite energy is necessary to change the vacuum solution of the scalar field in the whole space \cite{Rubakov:2002fi}. Our case is exactly the same –-- it is also necessary to have an infinite energy to change the vacuum solution with $h\phi_{0}>0$ in the whole space, whereas it is the corresponding massless mode that provides the attraction force forming Q-balls in the Wick--Cutkosky model.

The equations of motion, following from the action (\ref{sys}), take the form
\begin{eqnarray}
\Box\chi+h\phi\chi=0,\\
\Box\phi+h\chi^{*}\chi=0.
\end{eqnarray}
In the following, we will be looking for stationary spherically symmetric solutions without nodes, which take the form
\begin{eqnarray}\label{chiequiv}
\chi(t,\vec x)=e^{i\omega t}f(r),\\ \label{phiequiv}
\phi(t,\vec x)=\phi(r)
\end{eqnarray}
with $\partial_{r}f|_{r=0}=0$, $\lim\limits_{r\to\infty}f(r)=0$, $\partial_{r}\phi|_{r=0}=0$, $\lim\limits_{r\to\infty}\phi(r)=\phi_{0}$, where $r=\sqrt{{\vec x}^2}$ and $\omega$ is real.

For convenience, from the very beginning it is useful to represent the field $\phi$ as
\begin{equation}\label{vac-shift}
\phi=\phi_{0}+\tilde\phi
\end{equation}
and to consider the system of equations
\begin{eqnarray}\label{eq1}
-\omega^{2}f-\Delta f+m^2f+h\tilde\phi f=0,\\ \label{eq2}
-\Delta\tilde\phi+hf^2=0
\end{eqnarray}
with $m^2=h\phi_{0}>0$ and
\begin{eqnarray}\label{boundary1}
\partial_{r}f|_{r=0}=0,\qquad \lim\limits_{r\to\infty}f(r)=0,\\ \label{boundary2}
\partial_{r}\tilde\phi|_{r=0}=0, \qquad \lim\limits_{r\to\infty}\tilde\phi(r)=0.
\end{eqnarray}
Equations (\ref{eq1}) and (\ref{eq2}) follow from the action of the Wick--Cutkosky model in its common form in which the mass term of one of the scalar fields exists from the very beginning. In principle, the shift (\ref{vac-shift}) can be considered just as a redefinition of the field $\phi$ without any reference to vacuum, in this case one should consider the theory with the vacuum solution $\tilde\phi\equiv 0$.

In order to ensure that the field $f(r)$ falls off exponentially for $r\to\infty$, providing the finiteness of the Q-ball charge and energy, the frequency $\omega$ should be bounded as $|\omega|<m$ (moreover, it is not difficult to show that there are no Q-ball solutions for $\omega^{2}-m^{2}\ge 0$, including the case $m^{2}\le 0$; see Appendix~A). From the structure of these equations we expect that $\tilde\phi(r)\sim\frac{1}{r}$ for large $r$. We see that in the physically reasonable cases (i.e., in the cases with stable vacua) this system describes a massive charged scalar field in the long-range attractive potential provided by the field $\phi$.\footnote{The Schr\"{o}dinger--Poisson systems, which appear, for example, when one considers the Newtonian limit for boson stars made of scalar fields \cite{Ruffini:1969qy,Seidel:1990jh,Marsh:2015wka}, also provide equations of motion very similar to equations (\ref{eq1}) and (\ref{eq2}). But though the equations of motion in different models look similar from the mathematical point of view, the physical essence of different theories is completely different, starting from the origin of the coupling between the fields and ending with the definition of important physical characteristics of the solutions.}

\section{Q-ball solution and its properties}
\label{sec_charged}
Suppose that there exists a Q-ball solution to the system of equations (\ref{eq1}) and (\ref{eq2}), satisfying the boundary conditions (\ref{boundary1}) and (\ref{boundary2}). In this case, the $U(1)$ global charge of the Q-ball can be defined as
\begin{eqnarray}\label{chargedef}
Q=i\int\left(\chi\partial_{0}\chi^*-\chi^*\partial_{0}\chi\right)d^{3}x=4\pi\int\limits_{0}^{\infty}2\omega f^{2}r^{2}dr,
\end{eqnarray}
whereas the Q-ball energy takes the form
\begin{eqnarray}\label{energydef}
E=4\pi\int\limits_{0}^{\infty}\Bigl(\omega^{2}f^{2}+\partial_{r}f\partial_{r}f+m^{2}f^{2}+hf^{2}\tilde\phi+\frac{1}{2}\partial_{r}\tilde\phi\partial_{r}\tilde\phi\Bigr)r^{2}dr.
\end{eqnarray}
Using the equations of motion for the fields, it is not difficult to show that the Q-ball solution possesses the following properties:
\begin{eqnarray}
E&=&\omega Q+2\pi\int\limits_{0}^{\infty}(\partial_{r}\tilde\phi)^{2}r^{2}dr>\omega Q,\\ \label{dEdQ}
\frac{dE}{dQ}&=&\omega,
\end{eqnarray}
see also Appendix~B for details. The latter relation is well known for the one-field Q-balls.

The system of equations (\ref{eq1}) and (\ref{eq2}) can be brought into dimensionless form by means of the transformations
\begin{eqnarray}
R=r\sqrt{m^{2}-\omega^{2}},\qquad F(R)=\frac{h}{m^2-\omega^2}f(r),\qquad G(R)=\frac{h}{m^2-\omega^2}\tilde\phi(r),
\label{transformation}
\end{eqnarray}
resulting in
\begin{eqnarray}\label{eq1a}
-\Delta_{R}F+F+FG=0,\\ \label{eq2a}
-\Delta_{R}G+F^2=0
\end{eqnarray}
with the boundary conditions
\begin{eqnarray}\nonumber
\partial_{R}F|_{R=0}=0,\qquad \lim\limits_{R\to\infty}F(R)=0,\\ \label{boundary1a}
\partial_{R}G|_{R=0}=0, \qquad \lim\limits_{R\to\infty}G(R)=0.
\end{eqnarray}
Without loss of generality, we suppose that $F(R)>0$ for any $R$.

\begin{figure*}[!ht]
\centering
\includegraphics[width=0.95\linewidth]{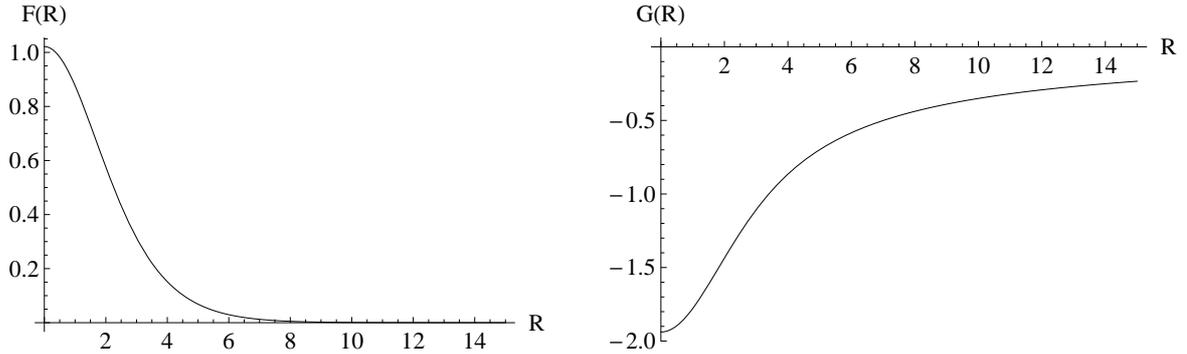}
\caption{Numerical solution for the functions $F(R)$ and $G(R)$.}\label{numsol}
\end{figure*}
In Fig.~\ref{numsol} the explicit numerical solution to equations (\ref{eq1a}), (\ref{eq2a}) with (\ref{boundary1a}) is presented (see Appendix~C for the details of the numerical analysis). In fact, this numerical solution is necessary only for the visualization of the Q-ball -- all the important characteristics of the Q-ball can be obtained either analytically or with the help of the auxiliary numerical solution presented in Appendix~C. The form of the solution implies that in general the thin-wall approximation cannot be used for describing solutions for the complex scalar field. Since the system of equations (\ref{eq1a}), (\ref{eq2a}) with (\ref{boundary1a}) is devoid of the parameters $\phi_{0}$, $h$, and $\omega$, the dimensionless solution presented in Fig.~\ref{numsol} can be used to restore the actual solution for the scalar fields $\chi$, $\phi$ for any physically reasonable choice of the coupling constant $h$, the vacuum expectation value $\phi_{0}$, and the frequency $\omega$.

We note that very often Q-ball solutions are obtained by minimization of the energy at a fixed charge. This procedure is fully equivalent to solving the corresponding equations of motion \cite{Coleman:1985ki,Friedberg:1976me}.

For the charge and the energy of the Q-ball we get
\begin{equation}\label{chargewickcutk}
Q=\frac{2\omega\sqrt{m^2-\omega^{2}}}{h^{2}}\,I, \qquad
E=\omega Q+\frac{(m^{2}-\omega^{2})^{\frac{3}{2}}}{h^{2}}\,J,
\end{equation}
where
\begin{equation}\label{Idef}
I=4\pi\int\limits_{0}^{\infty}F^{2}R^{2}dR,\qquad
J=2\pi\int\limits_{0}^{\infty}(\partial_{R}G)^{2}R^{2}dR.
\end{equation}
Using (\ref{dEdQ}), it is easy to show that $J=\frac{2}{3}I$, leading to
\begin{eqnarray}\label{Q}
Q&=&\frac{I}{h^2}2\,\omega\sqrt{m^2-\omega^{2}},\\ \label{E}
E&=&\frac{I}{h^2}\sqrt{m^2-\omega^{2}}\left(\frac{4}{3}\omega^{2}+\frac{2}{3}m^{2}\right).
\end{eqnarray}

\begin{figure}[!ht]
\centering
\includegraphics[width=0.8\linewidth]{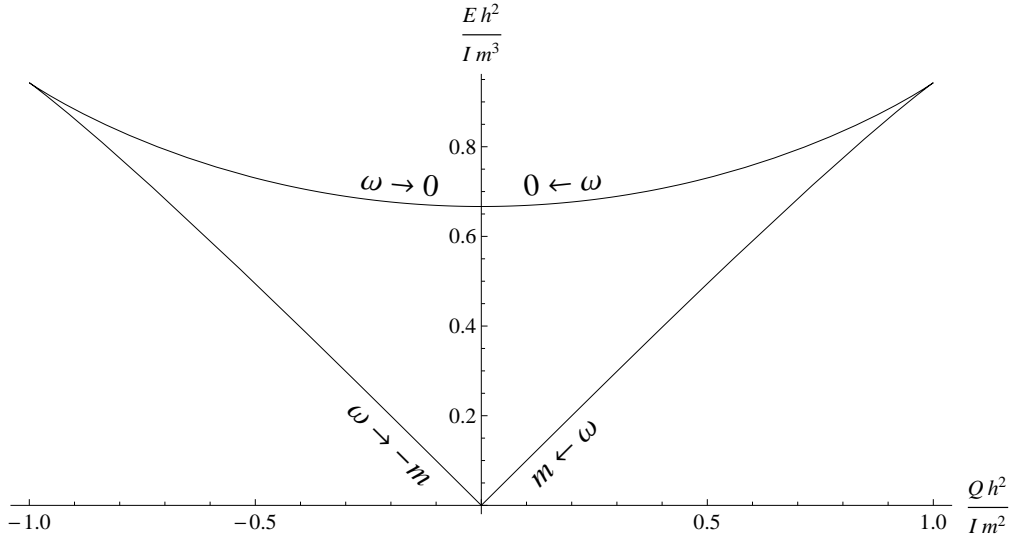}
\caption{$E(Q)$ dependence for Q-balls.}\label{figEQ}
\end{figure}

In Fig.~\ref{figEQ} the $E(Q)$ dependence, corresponding to (\ref{Q}), (\ref{E}), is presented.
The parameter $I$ can be found numerically and turns out to be $I\approx 44.05$ (see Appendix~C for details). The cusps in Fig.~\ref{figEQ}, corresponding to the points with $\frac{dQ}{d\omega}=0$ (at $\omega=-\frac{m}{\sqrt{2}}$ and $\omega=\frac{m}{\sqrt{2}}$), are a direct consequence of the relation (\ref{dEdQ}). The form of the $E(Q)$ dependence shows that there exist solutions with the maximal and minimal charges $Q_{max}=-Q_{min}=\frac{Im^{2}}{h^2}$, both possessing maximal energy $E_{max}=\frac{2\sqrt{2}\,Im^{3}}{3h^2}$. It should be mentioned that an analogous form of the $E(Q)$ dependence for Q-balls appears in the model of \cite{Rosen1}, describing a single complex scalar field with the logarithmic scalar field potential, see \cite{MarcVent}. However, contrary to the case of the Wick--Cutkosky model, in the model of \cite{Rosen1} the complex scalar field formally has an infinite mass.

From (\ref{Q}) and (\ref{E}) it is not difficult to obtain explicit formulas for this $E(Q)$ dependence: the upper branch in Fig.~\ref{figEQ} is described by
\begin{eqnarray}
E=\frac{\sqrt{2}\,I}{3\,h^{2}}\left(2m^{2}-\sqrt{m^{4}-\frac{Q^{2}h^{4}}{I^{2}}}\right)\sqrt{m^{2}+\sqrt{m^{4}-\frac{Q^{2}h^{4}}{I^{2}}}},
\end{eqnarray}
whereas the lower branches are described by
\begin{eqnarray}\label{energylowerbr}
E=\frac{\sqrt{2}\,I}{3\,h^{2}}\left(2m^{2}+\sqrt{m^{4}-\frac{Q^{2}h^{4}}{I^{2}}}\right)\sqrt{m^{2}-\sqrt{m^{4}-\frac{Q^{2}h^{4}}{I^{2}}}}.
\end{eqnarray}

Let us briefly discuss the stability of the obtained solutions. First, the stability criterion proposed in \cite{Friedberg:1976me,LeePang} implies that solutions on the lower branches of the $E(Q)$ dependence in Fig.~\ref{figEQ} are classically stable, i.e., stable with respect to small perturbations of the fields. Indeed, for these solutions the relation $\frac{dQ}{d\omega}<0$, which is one of the conditions necessary for the classical stability of such a two-field model \cite{Friedberg:1976me}, holds. Second, since $d^{2}E/dQ^{2}<0$ for the lower branches in Fig.~\ref{figEQ}, the Q-balls, corresponding to these solutions, are stable against fission (a simple justification of this fact in the general case can be found in \cite{Gulamov:2013ema}). And third, it is not difficult to check that the lower branches of the $E(Q)$ dependence in Fig.~\ref{figEQ} lie below the lines $E=m|Q|$, standing for free particles of mass $m$. The latter means that Q-balls, corresponding to these branches, are also quantum mechanically stable, i.e., stable with respect to decay into free scalar particles of mass $m$ (of course, the statement about the quantum mechanical stability is valid if there are no interactions with other particles in the theory under consideration). Thus, Q-ball solutions from the lower branches of the $E(Q)$ dependence in Fig.~\ref{figEQ} can be thought of as absolutely stable.

There exists a classically unstable time-independent solution with $\omega=0$ (and, consequently, $Q=0$) with nonzero finite energy. Such solutions are usually called ``sphalerons'' and can play an interesting role in quantum theory \cite{Manton:2004tk,Rubakov:2002fi}.

In this connection it is interesting to check whether the values of the initial scalar field $\phi$ (not $\tilde\phi$!) in the Q-ball are such that $h\phi$ becomes negative at least for some $r$, which corresponds to the area of unstable vacua in the scalar field potential. The smallest value of $h\phi$ is attained at $r=0$, so we will consider $h\phi(0)$. Using (\ref{vac-shift}), (\ref{transformation}) and the results presented in Appendix~C, we get
\begin{equation}
h\phi(0)=h\phi_{0}\left(1-C\left(1-\frac{\omega^{2}}{m^{2}}\right)\right),
\end{equation}
where we have used the fact that $h\phi_{0}=m^{2}$. Here $C\approx 1.938$; see Appendix~C. It is easy to find that $h\phi(0)\ge 0$ (and, consequently, $h\phi(r)\ge 0$ for any $r$) for $|\omega|\ge m\sqrt{\frac{C-1}{C}}\approx 0.696\cdot m$. We see that Q-balls from the lower (``stable'') branches (i.e., Q-balls with $m>|\omega|\ge \frac{m}{\sqrt{2}}\approx 0.707\cdot m$) and from the small parts of the upper (``unstable'') branch reside in the area of the scalar field potential (\ref{potential}) which corresponds to stable vacua, i.e., $V(\phi,\chi)>0$ for these Q-balls. This also indicates that there is no direct and simple connection between the values of the Q-ball scalar fields, the form of the scalar field potential, and the Q-ball stability, as noted in \cite{Nugaev:2013poa}. As for the rest of the Q-balls (for which $h\phi(0)<0$), some parts of such Q-balls reside in the area of the scalar field potential for which $V(\phi,\chi)<0$ holds, corresponding to unstable vacua. An interesting observation is that all Q-ball solutions with $h\phi(r)>0$ for any $r$ (which includes all Q-balls from the lower branches) can be reproduced in a theory with the scalar field potential
\begin{equation}
V(\phi,\chi)=h|\phi|\chi^*\chi,
\label{potential-modif}
\end{equation}
where $h>0$, which is bounded from below.

Now let us compare the Wick--Cutkosky model with the model proposed and examined in \cite{Levin:2010gp} (as noted in the Introduction, the latter is a simplification of the two-field model of \cite{Friedberg:1976me}). First, for Q-ball solutions in both models the real scalar field behaves as $\sim\frac{1}{r}$ at large $r$, so different Q-balls undergo long-range Coulomb attraction between each other regardless of the sign of their $U(1)$ global charges. As noted in \cite{Levin:2010gp}, one can say that Q-balls possess a (non-conserved) ``scalar charge'', which characterizes the strength of this long-range attraction. It will be defined explicitly in the next section.

Second, there is no sphaleron solution in the model of \cite{Levin:2010gp}. Indeed, according to the Derrick theorem \cite{Derrick}, there are no time-independent localized solutions in a theory with nonnegative scalar field potential, which is exactly the case of \cite{Levin:2010gp}.

Finally, it is possible to show that, for $M-|\omega|\ll M$, where $M$ is the mass of the free charged scalar particle ($M=m=\sqrt{h\phi_{0}}$ in the Wick--Cutkosky model), both models have almost the same $E(Q)$ dependencies such that $Q\to 0$, $E\to 0$ for $|\omega|\to M$; see Appendix~D for details. However, in the Wick--Cutkosky model the Q-ball energy is bounded from above, whereas in the model of \cite{Levin:2010gp} the charge and the energy go to infinity for $\omega\to 0$ such that $E\sim\sqrt{|Q|}$ for large $|Q|$. The latter implies that all Q-balls in the model of \cite{Levin:2010gp} can be absolutely stable (because $\frac{dQ}{d\omega}<0$ and $E<M|Q|$), whereas in our case there may exist absolutely stable and unstable Q-balls.

\section{Q-ball as a bound state of scalar particles}
As noted in the Introduction, Q-balls can be considered as bound states of scalar particles of the theory. In order to calculate the binding energy of a Q-ball, it is necessary to know the number of scalar particles forming the Q-ball. In this case the binding energy $\epsilon$ is simply
\begin{equation}\label{bindenergy}
\epsilon=Nm-E,
\end{equation}
where $N$ is the number of scalar particles in the Q-ball, $E$ is the Q-ball energy and $m$ is the mass of the free scalar particle of the theory. An obvious problem is to estimate the number of particles $N$.

Indeed, Q-balls are formed not only from particles, but also from anti-particles, i.e., it is a bound state of particles and anti-particles (it is obvious for sphalerons -- solutions with $Q=0$ and $E\neq 0$). In quantum field theory the operator of charge $\hat Q$ gives
\begin{equation}
\langle N_{+},N_{-}|\hat Q|N_{+},N_{-}\rangle=N_{+}-N_{-},
\end{equation}
where $|N_{+},N_{-}\rangle$ defines the state with $N_{+}$ particles and $N_{-}$ anti-particles, whereas we need $N=N_{+}+N_{-}$. However, it was shown in \cite{Levin:2010gp} that there exists a ``scalar charge'', which characterizes the strength of the long-range attraction of Q-balls in the model of \cite{Levin:2010gp} regardless of the sign of their global charges. The latter suggests that in the general case this ``scalar charge'' can be somehow connected with the total number of particles in a Q-ball.

In order to examine such a possibility, let us define the scalar charge as
\begin{equation}
Q_{SC}=2m\int\chi^*\chi\,d^{3}x.
\end{equation}
It is clear that this charge is {\em non-conserved} in general. But for Q-balls at rest we get
\begin{equation}\label{QscQball}
Q_{SC}=8\pi m\int\limits_{0}^{\infty}f^{2}r^{2}dr,
\end{equation}
which obviously is conserved in time. It is not difficult to check that for the non-relativistic particles, i.e., for $k_{0}-m\ll m$, the following relation holds in quantum field theory:
\begin{equation}
\langle N_{+},N_{-}|\hat Q_{SC}|N_{+},N_{-}\rangle\approx N_{+}+N_{-},
\end{equation}
where $\hat Q_{SC}=2m:\int\chi^*\chi\,d^{3}x:$, which is exactly what we need. So, we can assume that in a Q-ball $N\approx Q_{SC}$. Of course, this estimate can be used for $N\gg 1$.

\begin{figure}[!ht]
\begin{minipage}[t]{.49\linewidth}
\centering
\includegraphics[width=0.93\linewidth]{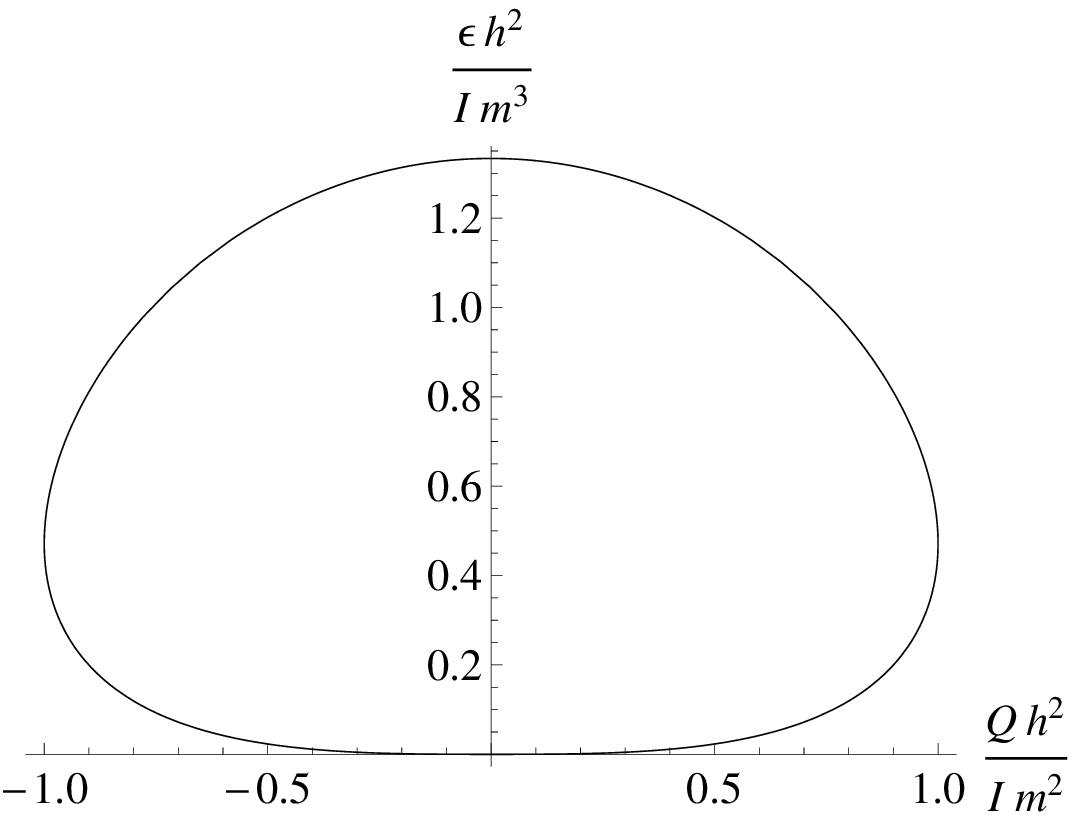}
\caption{$\epsilon(Q)$ dependence for Q-balls in the Wick--Cutkosky model.}\label{figEpsilonQ}
\end{minipage}
\begin{minipage}[t]{.49\linewidth}
\centering
\includegraphics[width=0.93\linewidth]{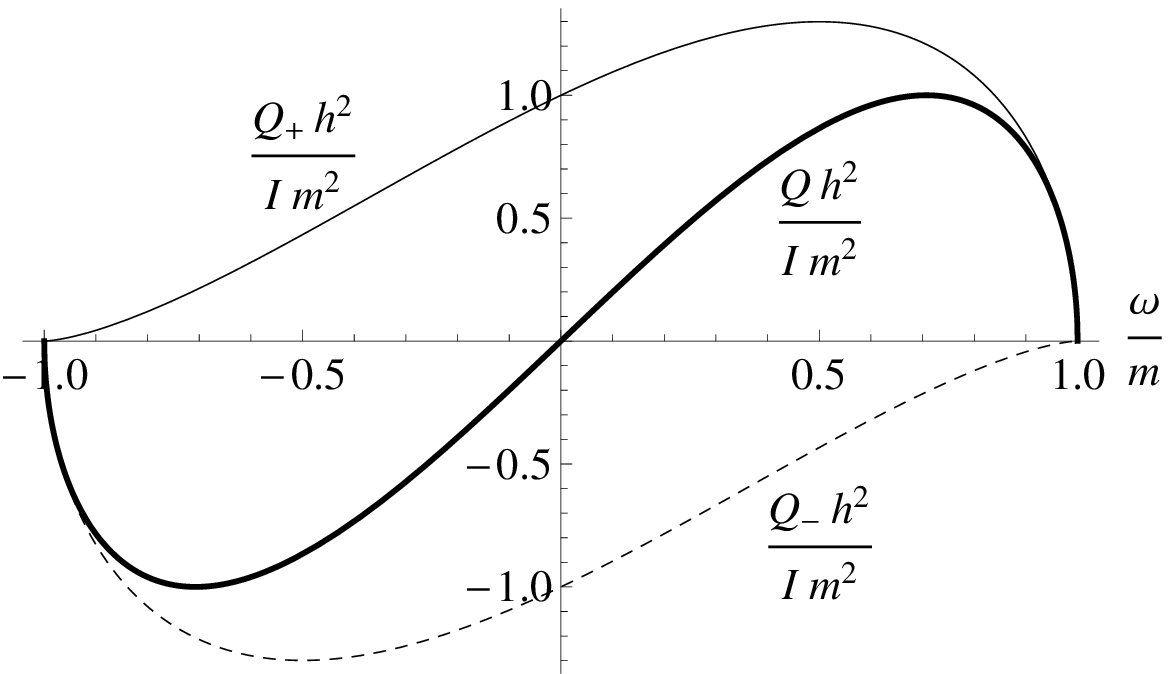}
\caption{$Q(\omega)$ (thick line), $Q_{+}(\omega)$ (thin line) and $Q_{-}(\omega)$ (dashed line) for Q-balls in the Wick--Cutkosky model.}\label{figQplusminus}
\end{minipage}
\end{figure}

Now, under the assumption that the most of scalar particles in the Q-ball are {\em non-relativistic}, with the help of (\ref{QscQball}) we can estimate the binding energy of any Q-ball. For example, for the Q-balls in the Wick--Cutkosky model we get from (\ref{QscQball}) and (\ref{bindenergy})
\begin{eqnarray}\label{schargewickcutk}
Q_{SC}(\omega)=\frac{2m\sqrt{m^2-\omega^{2}}}{h^{2}}\,I,\\ \label{bindingenergy} \epsilon(\omega)=\frac{4I}{3h^{2}}(m^{2}-\omega^{2})^{\frac{3}{2}}>0.
\end{eqnarray}
The corresponding $\epsilon(Q)$ dependence is presented in Fig.~\ref{figEpsilonQ}. It is also illustrative to consider the contributions of particles and anti-particles to the total charge $Q$:
\begin{equation}\label{nplusnminus}
Q_{+}=N_{+}\approx\frac{Q+Q_{SC}}{2},\quad Q_{-}=-N_{-}\approx\frac{Q-Q_{SC}}{2}.
\end{equation}
For the Q-balls in the Wick--Cutkosky model the corresponding plots are presented in Fig.~\ref{figQplusminus}.

Several comments are in order here. As noted above, in the scalar field theory there are no conserved charges giving $N$, $N_{+}$ or $N_{-}$, so we cannot define the quantities $N$, $N_{+}$ or $N_{-}$ in a completely rigorous way. However, at the classical level (when $N$ is supposed to be very large) we may assume that we have {\em approximately} $N$ particles in a Q-ball -- since there are no corresponding conserved charges, the ``actual'' numbers of particles and anti-particles can vary in time accounting for virtual processes of creation, annihilation, etc. inside a Q-ball. Meanwhile, the mean values $N_{+}$ and $N_{-}$ can have the following meaning -- if we add the energy $\epsilon$ to the Q-ball, we expect that it can disassemble into approximately $N_{+}$ free particles and $N_{-}$ free anti-particles (i.e., it is the binding energy of the whole Q-ball). It is the scalar charge $Q_{SC}$ that determines how Q-balls in the Wick--Cutkosky model (as well as in the model of \cite{Levin:2010gp}) interact with each other by means of the field $\tilde\phi$. The field $\tilde\phi$ is also responsible for forming a Q-ball from the quanta of the field $\chi$, so the choice of $Q_{SC}$ for determining the approximate number of particles and anti-particles makes sense. Although this choice is motivated by the properties of the Wick--Cutkosky model and the model of \cite{Levin:2010gp}, we think that this approach can be used in the general case too.

It is clear that unstable Q-balls cannot decay into $N=N_{+}+N_{-}$ particles, it is energetically forbidden. A Q-ball of charge $Q$ may decay into $\tilde N$ particles and/or anti-particles (with $|Q|\le\tilde N<N$) plus quanta of the field $\tilde\phi$. For example, a possible decay channel of sphalerons is the decay only into quanta of the massless field $\tilde\phi$. The latter is similar to the case of, say, positronium -- its mass is a few $eV$ less than twice the electron mass, and it decays only into photons. Thus, the description of a Q-ball as a collection of, say, $|Q|$ particles or anti-particles is inappropriate -- it obviously fails for sphalerons with $Q=0$, $E\neq 0$.

Now let us look at this problem from another point of view. Suppose that we have a Q-ball with, say, charge $Q>0$ and we add a single particle of mass $m$ (and charge $+1$) to it. If $Q\gg 1$ (i.e., if $\frac{Im^{2}}{h^2}\gg 1$), then, according to (\ref{dEdQ}), we get the Q-ball of charge $Q+1$ and with the energy $E(Q+1)\approx E(Q)+\omega$ (here $\omega$ is similar to the chemical potential in thermodynamics). It is clear that in order to release a particle from this Q-ball, we should add the energy $m-\omega$ to the Q-ball.\footnote{When we add a particle to the Q-ball, the remaining energy $m-\omega$ should be radiated out, say, by the field $\phi$.} For the anti-particle, the corresponding ``chemical potential'' is $-\omega$, whereas the ``ionization'' energy is $m+\omega$. One may think that by adding a particle to a Q-ball, we just increase the corresponding number $N_{+}$ by unity. It is not so in general. For example, for $\omega=\frac{m}{2}$, using (\ref{Q}), (\ref{schargewickcutk}) and (\ref{nplusnminus}) we get $\Delta N_{+}\approx 0$, $\Delta N_{-}\approx -1$. In this case the interpretation is clear -- namely, the incoming particle annihilate with an anti-particle inside the Q-ball. However, for $\omega=0$, we get $\Delta N_{+}\approx\frac{1}{2}$, $\Delta N_{-}\approx -\frac{1}{2}$, which has no such a clear interpretation. For $\omega\to m$ we get $\Delta N_{+}\approx 1$, $\Delta N_{-}\approx 0$. These examples indicate that the numbers $N_{+}$, $N_{-}$, which are specific to each Q-ball, should be considered only as approximate values for estimates. Namely, if we increase the charge of a Q-ball by a finite value $\Delta Q>0$, the values $\Delta N_{+}=N_{+}(Q+\Delta Q)-N_{+}(Q)$ and $\Delta N_{-}=N_{-}(Q+\Delta Q)-N_{-}(Q)$ indicate, of course approximately, how many of the incoming particles are absorbed by the Q-ball and how many anti-particles inside the Q-ball are annihilated by the incoming particles. A fully analogous considerations can be made for $\Delta Q<0$, i.e., for the incoming anti-particles.

Finally, let us compare the Q-ball binding energy, obtained using the procedure presented in this section, with the energy of the bound state coming from the solution to the Bethe--Salpeter equation for $Q=2$, $\frac{h^{2}}{Im^{2}}\ll 1$. For the comparison, one should take Q-balls from the lower branches of Fig.~\ref{figEQ}, for which $N\approx 2$. Of course, Q-ball is a classical object and the use of our classical solution is not justified for such small values of $Q$ and $E$. However, an analogous comparison of the energy of a small Q-ball in the Minimal Supersymmetric Standard Model with the energy of the two-particle bound state in the Wick--Cutkosky model, which was performed in \cite{Kusenko:1998yj}, revealed a good qualitative agreement. Here we have the Q-ball solution in the Wick--Cutkosky model, so such a comparison can be even more illustrative.

The energy of the bound state in the Wick--Cutkosky model in the leading order in $\frac{h^{4}}{m^{4}}$ takes the form
\begin{equation}\label{BSWC}
E_{BS}=m\left(2-\left(\frac{h^{2}}{16\pi m^{2}}\right)^{2}\frac{1}{4n^{2}}\right),
\end{equation}
where $n=1,2,3,...$; see, for example, \cite{Feldman:1973jg,Darewych:1998mb}. As for the Q-balls, since the use of classical solutions is not justified for such small values of $Q$ and it is not clear what quantity should exactly correspond to $E_{BS}$, we will take both (\ref{energylowerbr}) and (\ref{bindingenergy}). Thus, in the leading order in $\frac{h^{4}}{m^{4}}$ and for $Q=2$ we get from (\ref{energylowerbr}) and (\ref{Q}), (\ref{bindingenergy})
\begin{eqnarray}
E&\approx&m\left(2-\frac{h^{4}}{3I^{2}m^{4}}\right),\\
\epsilon&\approx&m\frac{4h^{4}}{3I^{2}m^{4}}.
\end{eqnarray}
Using $I\approx 44.05$ and with $n=1$ in (\ref{BSWC}), we obtain
\begin{eqnarray}
2m-E_{BS}&\approx& \frac{h^{4}}{m^{3}}\,9.9\times 10^{-5},\\
2m-E&\approx& \frac{h^{4}}{m^{3}}\,1.7\times 10^{-4},\\
\epsilon&\approx& \frac{h^{4}}{m^{3}}\,6.9\times 10^{-4}.
\end{eqnarray}
Surprisingly, the agreement is very good taking into account the use of a classical solution for describing purely quantum effects. This implies that, for $N\gg 1$, for which the use of classical Q-ball solutions is justified, the method presented above indeed can give a result somewhat close to the real binding energy and it can be used at least for rough estimates.

\section*{Acknowledgements}
The authors are grateful to S.~Troitsky for valuable discussions. The work was supported by the Grant 16-12-10494 of the Russian Science Foundation.

\section*{Appendix~A: The absence of Q-balls for $\omega^{2}-m^{2}\ge 0$}
According to (\ref{chiequiv})--(\ref{vac-shift}), we can use the effective action
\begin{eqnarray}\label{effactapp}
S_{eff}=\int\Bigl((\omega^{2}-m^{2})f^{2}-\partial_{i}f\partial_{i}f-\frac{1}{2}\partial_{i}\tilde\phi\partial_{i}\tilde\phi-h\tilde\phi f^{2}\Bigr)d^3x,
\end{eqnarray}
where $i=1,2,3$, instead of the initial action. Suppose that there exists a Q-ball solution $f(\vec x)$, $\tilde\phi(\vec x)$. Let us apply the scale transformation $f(\vec x)\to f_{\lambda}(\vec x)=\lambda f(\lambda\vec x)$, $\tilde\phi(\vec x)\to \tilde\phi_{\lambda}(\vec x)=\lambda \tilde\phi(\lambda\vec x)$ to this solution and substitute the result into the effective action (\ref{effactapp}) instead of the original solution (in fact, it is just a generalization of the technique which was used in \cite{Derrick} to show the absence of time-independent soliton solutions in some nonlinear scalar field theories). We get
\begin{eqnarray}\nonumber
S_{eff}^{\lambda}=\int d^3y\frac{1}{\lambda^{3}}\Biggl(\lambda^{2}(\omega^{2}-m^{2})f^{2}(\vec y)-\lambda^{4}\frac{\partial f(\vec y)}{\partial y^{i}}\frac{\partial f(\vec y)}{\partial y^{i}}\\
-
\lambda^{4}\frac{1}{2}\frac{\partial \tilde\phi(\vec y)}{\partial y^{i}}\frac{\partial \tilde\phi(\vec y)}{\partial y^{i}}-\lambda^{3}h\tilde\phi(\vec y) f^{2}(\vec y)\Biggr),
\end{eqnarray}
where we have passed to the new coordinates $\vec y=\lambda \vec x$. According to the principle of least action
\begin{equation}
\frac{S_{eff}^{\lambda}}{d\lambda}\biggl|_{\lambda=1}=0,
\end{equation}
which results in
\begin{eqnarray}\label{scaletr}
(\omega^{2}-m^{2})\int f^{2}\,d^3x+\int\partial_{i}f\partial_{i}f\,d^3x+\frac{1}{2}\int\partial_{i}\tilde\phi\partial_{i}\tilde\phi\,d^3x=0.
\end{eqnarray}
It is clear that if $\omega^{2}-m^{2}\ge 0$ (which includes the case $m^{2}=h\phi_{0}\le 0$), then the only solution to (\ref{scaletr}) is
\begin{eqnarray}
f(\vec x)\equiv 0,\\ \tilde\phi(\vec x)\equiv 0.
\end{eqnarray}

\section*{Appendix~B: The relation $\frac{dE}{dQ}=\omega$}
It is reasonable to suppose that the only parameter, which
characterizes the charge and the energy of the Q-ball, is $\omega$. Thus, differentiating (\ref{energydef}) with respect to $\omega$, we get
\begin{eqnarray}\nonumber
\frac{dE}{d\omega}=4\pi\int\left(2\omega f^2+2\omega^2f\frac{df}{d\omega}+
2\partial_{r}f\partial_{r}\frac{df}{d\omega}+2m^{2}f\frac{df}{d\omega}\right.\\
\nonumber+\left.2hf\frac{df}{d\omega}\tilde\phi+hf^{2}\frac{d\tilde\phi}{d\omega}+\partial_{r}\tilde\phi\partial_{r}\frac{d\tilde\phi}{d\omega}
\right)r^{2}dr.
\end{eqnarray}
Integrating by parts the terms with derivatives in the latter formula (since it is supposed that $\tilde\phi(r)\sim\frac{1}{r}$ for $r\to\infty$ and consequently $\frac{d\tilde\phi(r)}{d\omega}\sim\frac{1}{r}$, the surface term, arising when an integration by parts is
performed, obviously vanishes), using equations of motion (\ref{eq1}), (\ref{eq2}) and the definition of the charge (\ref{chargedef}),
we get
\begin{eqnarray}\nonumber
\frac{dE}{d\omega}=4\pi\int\left(2\omega f^2+4\omega^2f\frac{df}{d\omega}\right)r^{2}dr=\omega\frac{dQ}{d\omega},
\end{eqnarray}
leading to
\begin{eqnarray}\label{dEdQgauged}
\frac{dE}{dQ}=\omega
\end{eqnarray}
for $\frac{dQ}{d\omega}\ne 0$. The points at which $\frac{dQ}{d\omega}=0$
(and, consequently, $\frac{dE}{d\omega}=0$) correspond to the cusps on the $E(Q)$ diagram and separate ``stable'' and ``unstable'' branches.

\section*{Appendix~C: Numerical solution}
In order to solve the system of equations (\ref{eq1a}), (\ref{eq2a}) numerically, it is convenient to pass to the new variables
\begin{eqnarray}\label{auxsol1}
F(R)=(C-1)\hat F(Y),\\ \label{auxsol2}
G(R)=-C+(C-1)\,\hat G(Y),\\
Y=\sqrt{C-1}\,R,
\end{eqnarray}
where $C=-G(0)>1$. In these notations the system of equations (\ref{eq1a}), (\ref{eq2a}) can be rewritten as
\begin{eqnarray}\label{eq1b}
-\Delta_{Y}\hat F-\hat F+\hat F\hat G=0,\\ \label{eq2b}
-\Delta_{Y}\hat G+{\hat F}^2=0
\end{eqnarray}
with the boundary conditions
\begin{eqnarray}\nonumber
\partial_{Y}\hat F|_{Y=0}=0,\qquad \lim\limits_{Y\to\infty}\hat F(Y)=0,\\ \label{boundary1b}
\partial_{Y}\hat G|_{Y=0}=0, \qquad \hat G|_{Y=0}=0.
\end{eqnarray}
Without loss of generality, we suppose that $\hat F(Y)>0$ for any $Y$.

\begin{figure*}[!h]
\centering
\includegraphics[width=0.95\linewidth]{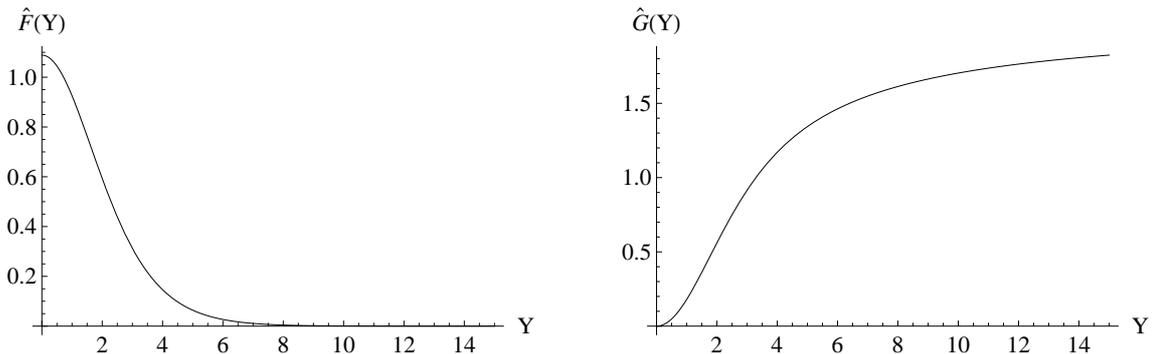}
\caption{Numerical solution for the functions $\hat F(Y)$ and $\hat G(Y)$.}\label{numsolaux}
\end{figure*}

Contrary to the case of initial equations (\ref{eq1a}), (\ref{eq2a}) with (\ref{boundary1a}), in which it is necessary to scan over the two parameters $F(0)$ and $G(0)$ in order to find a solution, in the case of the system of equations (\ref{eq1b}), (\ref{eq2b}) with (\ref{boundary1b}) one should scan over only the one parameter $\hat F(0)$ searching for such $\hat F(Y)$ that it falls off exponentially for $Y\to\infty$. The corresponding solution to equations (\ref{eq1b}), (\ref{eq2b}) with (\ref{boundary1b}) can easily be found numerically; the result is presented in Fig.~\ref{numsolaux}.
The value of $\hat F(0)$ for this solution turns out to be $\hat F(0)\approx 1.089$.

Now let us calculate the value of the constant $C$. From equation (\ref{eq2b}) it follows that
\begin{equation}
\hat G(\infty)=\int\limits_{0}^{\infty}{\hat F}^2(Y)YdY.
\end{equation}
On the other hand, from (\ref{auxsol2}) we get
\begin{equation}
C=(C-1)\,\hat G(\infty),
\end{equation}
leading to
\begin{equation}
C=\frac{\hat G(\infty)}{\hat G(\infty)-1}=\frac{\int_{0}^{\infty}{\hat F}^2(Y)YdY}{\int_{0}^{\infty}{\hat F}^2(Y)YdY-1}.
\end{equation}
Using the numerical solution for $\hat F(Y)$, the value of the constant $C$ was found to be $C\approx 1.938$. Using the explicit value of the constant $C$, it is not difficult to restore the solution to equations (\ref{eq1a}), (\ref{eq2a}) with (\ref{boundary1a}), which is presented in Fig.~\ref{numsol}, from the numerical solution for the functions $\hat F(Y)$ and $\hat G(Y)$.

The last step is to find the value of the parameter $I$ in formulas (\ref{Q}), (\ref{E}). According to the definition (\ref{Idef}) of the parameter $I$ and with the help of (\ref{auxsol1}), we get
\begin{equation}
I=4\pi\sqrt{C-1}\int\limits_{0}^{\infty}{\hat F}^{2}(Y)Y^{2}dY.
\end{equation}
The latter integral can also easily be evaluated numerically, resulting in $I\approx 44.05$.

\section*{Appendix~D: The model of \cite{Levin:2010gp} for $M-|\omega|\ll M$}
The scalar field potential in \cite{Levin:2010gp} has the form $h\phi^{2}\chi^*\chi$. For the vacuum solution $\chi\equiv 0$, $\phi\equiv\phi_{0}$ the mass of the free charged scalar particle is defined by $M^{2}=h\phi_{0}^{2}$. Using (\ref{chiequiv}), (\ref{phiequiv}) and the redefinition
\begin{eqnarray}
R=r\sqrt{M^{2}-\omega^{2}},\qquad F(R)=\frac{2h\phi_{0}}{M^2-\omega^2}f(r),\qquad G(R)=\frac{2h\phi_{0}}{M^2-\omega^2}\tilde\phi(r),
\end{eqnarray}
we get for the model of \cite{Levin:2010gp}
\begin{eqnarray}\label{eqLRa}
-\Delta_{R}F+F+FG+\frac{1}{4}\lambda(\omega) FG^{2}=0,\\ \label{eqLRb}
-\Delta_{R}G+F^2+\frac{1}{2}\lambda(\omega) F^{2}G=0,
\end{eqnarray}
where $\lambda(\omega)=1-\frac{\omega^{2}}{M^{2}}$. It is clear that $\lambda(\omega)\ll 1$ for $M-|\omega|\ll M$ and the terms proportional to $\lambda(\omega)$ in (\ref{eqLRa}), (\ref{eqLRb}) can be neglected in this case, leading to the system of equations (\ref{eq1a}), (\ref{eq2a}). Thus, the $E(Q)$ dependence for Q-balls in the model of \cite{Levin:2010gp} for very small charges is almost the same as the one for Q-balls with small charges from the lower branches of Fig.~\ref{figEQ}.

\end{document}